نوع مقاله: پژوهشی

# تحلیل کمی چالش‌های پیاده‌سازی زنجیره‌تأمین دیجیتال مبتنی بر فناوری اینترنت اشیا (زنجیره‌تأمین ۴.۰)

حامد نوذری*¹، محمدابراهیم صادقی²، جاوید قهرمانی نهر³، سیداسماعیل نجفی⁴

۱. استادیار، گروه مهندسی صنایع دانشگاه آزاد اسلامی واحد تهران مرکزی، تهران، ایران.

۲. گروه مدیریت صنعتی، دانشکده مدیریت، دانشگاه تهران، تهران، ایران.

۳. دانشجوی دکترا، پژوهشکده توسعه برنامه‌ریزی جهاد دانشگاهی، تبریز، ایران.

۴. دانشیار، گروه مهندسی صنایع دانشگاه آزاد اسلامی واحد علوم و تحقیقات تهران، تهران، ایران.

| سابقه مقاله | چکیده |
|---|---|

در طول سی سال گذشته، لجستیک دستخوش تغییر عظیمی شده است و از یک عملکرد صرفاً عملیاتی که به فروش یا تولید منتهی می‌شد و بر تضمین تأمین در خطوط تولید و تحویل به مشتریان متمرکز بود، با همراهی فناوری‌های تحول‌آفرین و هوشمند به یک عملگر حرفه‌ای و مستقل در مدیریت زنجیره‌تأمین تبدیل شده است. در دنیای امروزی نسل چهارم صنعتی شرکت‌ها را ملزم کرده است تا در نحوه طراحی زنجیره‌تأمین خود تجدیدنظر کنند. در این حالت، علاوه بر نیاز به انطباق، زنجیره‌های تأمین همچنین این فرصت را دارند که به افق اثربخشی عملیاتی برسند، از مدل‌های تجاری زنجیره‌تأمین دیجیتال در حال ظهور استفاده کنند و شرکت را به زنجیره‌تأمین دیجیتال تبدیل کنند. یکی از فناوری‌های تحول‌آفرین که در این راستا تأثیر شگرفی بر زنجیره تأمین داشته است، فناوری اینترنت اشیا می‌باشد. این فناوری به‌عنوان یکی از بزرگ‌ترین منابع تولید داده‌ها می‌توانند تسهیل‌کننده فرایندهای زنجیره‌تأمین در همه ابعاد آن باشد. اما به‌واسطه حضور اینترنت و قرارگرفتن اجزای زنجیره‌تأمین در بستر شبکه‌های اطلاعاتی، همواره این زنجیره‌تأمین دیجیتال با چالش‌های اساسی روبه‌رو می‌باشد. ازاین‌رو در این مقاله کوشیده شد تا مهم‌ترین چالش‌های پیاده‌سازی یک زنجیره‌تأمین ۴.۰ با استفاده از یک روش تحلیل سلسله‌مراتبی غیرخطی مورد بررسی و اولویت‌بندی قرار گیرد. . به‌منظور بررسی این چالش‌ها در زنجیره تأمین صنایع تند مصرف (FMCG) به‌عنوان مطالعه موردی و همین‌طور برخی از متخصصان دانشگاهی بهره گرفته شد. نتایج نشان داده که عدم وجود زیرساخت‌های تکنولوژیکی و چالش‌های امنیتی از مهم‌ترین چالش‌های پیاده‌سازی زنجیره‌تأمین ۴.۰ در عصر تحولات دیجیتالی می‌باشند که باید برای یک پیاده‌سازی موفق توجه ویژه‌ای به بدان‌ها

**کلمات کلیدی:** زنجیره‌تأمین دیجیتال، چالش‌های پیاده‌سازی زنجیره‌تأمین دیجیتال، زنجیره‌تأمین مبتنی بر اینترنت اشیا، زنجیره‌تأمین ۴.۰.

# ۱  مقدمه

مدیریت زنجیره‌تأمین شامل برنامه‌ریزی و مدیریت تمام فعالیت‌های مرتبط با تأمین منابع و تدارکات و تمامی فعالیت‌های مدیریت لجستیک است. مهم‌تر از همه، این امر شامل هماهنگی و همکاری با شرکای کانال‌های مختلف است که می‌توانند تأمین‌کنندگان، واسطه‌ها، ارائه‌دهندگان خدمات شخص ثالث و مشتریان باشند. در اصل، مدیریت زنجیره‌تأمین، مدیریت عرضه و تقاضا را در داخل و بین شرکت‌ها یکپارچه می‌کند. مدیریت زنجیره‌تأمین یک عملکرد یکپارچه با مسئولیت اصلی برای پیوند دادن عملکردهای تجاری اصلی و فرایندهای تجاری در داخل و بین شرکت‌ها به یک مدل تجاری منسجم و با عملکرد بالا است. این شامل تمام فعالیت‌های مدیریت لجستیک ذکر شده در بالا و همچنین عملیات تولیدی است و هماهنگی فرایندها و فعالیت‌ها با بازاریابی، فروش، طراحی محصول، امور مالی و فناوری اطلاعات را به همراه دارد (نوذری و همکاران، ۲۰۱۹).

زنجیره‌تأمین دیجیتال به‌صورت متداول برای بحث در مورد چگونگی توسعه و اجرا و به‌کارگیری فناوری‌های دیجیتال پیشرفته (اینترنت اشیا، بلاک‌چین، یادگیری ماشین، هوش‌مصنوعی، تجزیه‌وتحلیل پیش‌بینی‌کننده و غیره) در فرایندهای زنجیره‌تأمین استفاده می‌شود (قهرمانی نهر و همکاران، ۲۰۲۱). به‌عبارت‌دیگر "مدیریت زنجیره‌تأمین دیجیتال" در واقع فقط مدیریت زنجیره‌تأمین با لایه‌ای از فناوری‌های دیجیتال است که به بدان اضافه شده است. این فناوری‌ها به تحلیل و پیش‌بینی برای بهینه‌سازی تخصیص موجودی و پیش‌بینی تقاضا و همین‌طور راه‌حل‌های تکمیل خودکار برای سرعت بخشیدن به مونتاژ یا چیدمان در زنجیره‌تأمین کمک می‌کنند. این زنجیره‌تأمین هوشمند شده با استفاده از فناوری‌های تحول‌آفرین مانند هوش‌مصنوعی و اینترنت اشیا، زنجیره‌تأمین ۴٫۰ نامیده می‌شود. بدیهی است که چرا برای شرکت‌ها این‌قدر مهم است که زنجیره‌تأمین دیجیتال را به‌درستی پیاده‌سازی کنند: وقتی یک مدل زنجیره‌تأمین قوی توسط بهترین فناوری‌ها (مانند اینترنت اشیا، هوش مصنوع یو یادگیری ماشین) پشتیبانی می‌شود، شرکت‌ها را در برابر تغییر در زنجیره‌تأمین انعطاف‌پذیر می‌کند. این امر به انعطاف‌پذیری کلی سازمان کمک می‌کند، که در شرایط اقتصادی نامطمئن، مانند شرایطی که در حال حاضر با آن

روبرو هستیم، حیاتی است. همچنین شرکت‌ها را قادر می‌سازد تا برای پاسخگویی به شرایط مختلف بازار، انعطاف‌پذیر و سازگار باشند و از پتانسیل رشد و فرصت‌ها برای تولیدکنندگان و زنجیره‌تأمین به‌عنوان یک کل حمایت کنند (نوذری و همکاران، ۲۰۲۱).

در عصر تحولات دیجیتالی سیستم‌های برنامه‌ریزی منابع سازمانی، مدل‌های زنجیره‌تأمین را قادر می‌سازد تا قوی‌تر کار کنند. گرایش به سمت سرویس‌دهی که در آن کسب و کارها از طریق خدمات اضافی، مانند مراقبت‌های بیشتر، به ارزش ارائه محصول خود می‌افزایند، با پیوند دادن کارکردهای تجاری لازم در زنجیره امکان‌پذیر شده است. روندهای جدیدتر - از جمله اقتصاد دایره‌ای که بر بازسازی و بازیافت مواد متمرکز شده است - با اشتراک‌گذاری داده‌های کارآمد در سراسر شبکه زنجیره‌تأمین افزایش شدید کارایی را به همراه دارد. زنجیره‌های تأمین دیجیتال به روشی محبوب برای تقویت زنجیره‌های تأمین سنتی تبدیل شده‌اند. در سطح پایه، تفاوت‌های بسیار کمی بین یک زنجیره‌تأمین سنتی و یک زنجیره‌تأمین دیجیتال وجود دارد. هر دو شامل تأمین منابع، تبدیل مواد به محصول، توزیع محصول و سپس فروش محصول است. با این حال، زنجیره‌های تأمین دیجیتال می‌توانند این وظایف را از طریق فرآیندهای دیجیتالی خودکار ردیابی و تکمیل کنند که کارایی بیشتری را ممکن می‌سازد. تجزیه‌وتحلیل پیش‌بینی‌کننده در برنامه‌ریزی تقاضا صدها تا هزاران متغیر تأثیرگذار بر تقاضای داخلی و خارجی (مانند آب‌وهوا، روندهای شبکه‌های اجتماعی، داده‌های حسگر) را با رویکردهای شبکه بیزی و یادگیری ماشین برای کشف و مدل‌سازی روابط پیچیده و استخراج دقیق تجزیه و تحلیل می‌کند (بنرجی[1] و همکاران، ۲۰۲۱). یک زنجیره‌تأمین دیجیتال مؤثر به سازمان‌ها کمک می‌کند تا با افزایش تقاضای کسب‌وکار همراه باشند – اما بسیاری از آن‌ها در ایجاد و بهینه‌سازی زنجیره‌تأمین دیجیتال خود با مشکل مواجه هستند. برخی محققان نشان دادند که شرکت‌هایی که اکثر زنجیره‌تأمین خود را دیجیتالی می‌کنند، می‌توانند درآمد خود را ۳٫۲ درصد و رشد درآمد سالانه

---

خود را به‌طور متوسط ۲٫۳ درصد افزایش دهند. با زنجیره‌تأمین دیجیتال، سازمان‌ها قادر است از افزایش درآمد، بهبود تصمیم‌گیری و فرایندهای چابک‌تر لذت ببرد. ولی متأسفانه، سازمان‌ها معمولاً هنگامی که دیجیتالی کردن زنجیره‌تأمین خود را آغاز می‌کنند، با چالش‌های مختلفی مواجه می‌شوند. در این مقاله کوشیده شده است که مهم‌ترین چالش‌های اساسی یک زنجیره‌تأمین داخلی با تاکید بر مرور ادبیات و بررسی نظرات خبرگان استخراج و اولویت‌بندی شود همچنین در این مطالعه از نظرات خبرگان فعال در صنایع تند مصرف (شرکت‌های دارویی و غذایی) به دلیل دسترسی مؤلفان به این شرکت‌ها و همین‌طور آشنایی این متخصصان با فناوری‌های تحول‌آفرین به‌عنوان مطالعه موردی بهره گرفته شده است. برای تحلیل این موضوع از یک روش تحلیل سلسله مراتبی غیرخطی (میخایلوف) استفاده شده است. درک درست و اولویت صحیح چالش‌ها می‌تواند به پیاده‌سازی مؤثر یک زنجیره‌تأمین دیجیتال و افزایش رشد و در عین حال بهره‌وری سازمانی و تسهیل فعالیت‌ها کمک نماید. همچنین بررسی و اولویت‌بندی این چالش‌ها مخصوصا در صنایع تند مصرف (به واسطه نوع و اهمیت بالای محصولات و طیف عظیم مخاطبان) می‌تواند به سازمان‌ها کمک نماید تا با رفع چالش‌های پر اهمیت‌تر، از بیشینه امکانات و تسهیلات زنجیره‌تأمین دیجیتال استفاده نموده و چابکی و تاب‌آوری زنجیره‌تأمین را در سازمان افزایش دهند.

ساختار این مقاله به این صورت تدوین شده است. در بخش دوم مرور ادبیات مرتبط با زنجیره‌تأمین دیجیتال و زنجیره‌تأمین ۴٫۰ مورد واکاوی قرار گرفته است. در بخش سوم چالش‌های اساسی پیاده‌سازی یک زنجیره‌تأمین مبتنی بر فناوری‌های دیجیتال ارائه می‌گردد. در بخش چهارم روش پژوهش مطرح می‌شود. در بخش پنجم مدل کمی ایجاد شده و یافته‌های پژوهش ارائه می‌گردد و در نهایت در بخش ششم نتیجه‌گیری بیان خواهد شد.

## ۲  پیشینه تحقیق

در دهه اخیر فناوری اطلاعات و انواع فناوری‌های تحول‌آفرین مانند هوش‌مصنوعی و یادگیری ماشین، تحلیلی داده‌های بزرگ و علوم داده‌ها، اینترنت اشیا و سایر فناوری‌ها، تأثیر بسیار زیادی بر

موقعیت و عملکرد بسیاری از بخش‌های جوامع بشری، سازمان‌ها و افراد جامعه گذاشته و تغییرات عظیمی در این زمینه بوجود آمده است.درک درست این فناوری‌ها و بلوغ سازمانی برای استفاده کامل از این فناوری‌ها در راستای هوشمندسازی همواره می‌تواند مزایای بسیاری برای سازمان‌ها به همراه داشته باشد. ولی به دلیل حضور شبکه و به هم پیوستگی که این فناوری‌ها همواره به همراه خود دارند، بنابراین پیاده‌سازی درست این فناوری‌ها و هوشمند کردن سازمان‌ها همواره می‌تواند با چالش‌هایی نیز همراه باشد. درک درست این چالش‌ها خود می‌تواند کمک کند تا بتوانیم با رفع آن‌ها از مزایای حداکثری این این فناوری‌های هوشمندساز استفاده نموده و تسهیلات بیشتری را برای زندگی خود و اطرافیان فراهم آوریم. در سال‌های اخیر تحقیقات مختلفی در خصوص بررسی چالش‌های پیاده‌سازی سیستم‌های هوشمند صورت گرفته است که خود نشان از اهمیت بررسی چالش‌ها و جذاب بودن این مفاهیم دارد. به‌عنوان مثال سیلوا[1] و همکاران (۲۰۱۸) در پژوهشی به بررسی چالش‌های باز در پیاده‌سازی شهرهای هوشمند پرداختند. لیم[2] و همکاران (۲۰۱۸) نیز در پژوهشی به چالش‌های مبتنی بر داده‌های عظیم در شهرهای هوشمند دیجیتال پرداخته‌اند و مدلی نیز برای پیاده‌سازی ارائه دادند. چن[3] و همکاران (۲۰۱۷) ابعاد و مؤلفه‌ها و چالش‌های پیاده‌سازی یک کارخانه هوشمند دیجیتال را مورد بررسی و واکاوی قرار دادند. پیرو و زامفرسکو[4] (۲۰۱۷) چالش‌ها و فرصت‌های پیاده‌سازی یک کارخانه هوشمند دیجیتال را در کشور رومانی مورد بحث قرار داده و شاخص‌های کلیدی را بر اساس نظرات خبرگان در این کشور استخراج کردند. یی و ژورنگ[5] (۲۰۲۱) به بررسی چالش‌های امنیتی مرتبط با پیاده‌سازی کارخانه‌های هوشمند پرداختند و اهمیت این چالش‌ها را نیز بررسی کردند. در خصوص بررسی چالش‌های زنجیره‌تأمین مبتنی بر فناوری‌های تحول‌آفرین نیز برخی پژوهش‌ها در طی سالیان اخیر صورت گرفته است. در این راستا صلاح[6] و همکاران (۲۰۱۹) چالش‌های حضور فناوری بلاک چین در زنجیره‌تأمین دیجیتال را بررسی

---

1 Silva
2 Lim
3 Chen
4 Pîrvu & Zamfirescu
5 Yi & Jeong
6 Salah

کردند. شاو[1] و همکاران (۲۰۲۱) نیز با بررسی چالش‌های پیش روی یک زنجیره‌تأمین دیجیتال در عصر فناوری‌های نوین، چارچوبی را برای پیاده‌سازی یک زنجیره‌تأمین قدرتمند دیجیتال ارائه دادند. اسادی[2] (۲۰۲۱) نیز با توسعه زنجیره‌تأمین دیجیتال در یک بیمارستان هوشمند به بررسی چالش‌های پیاده‌سازی آن در سیستم‌های درمانی پرداخته است. بنابراین به‌نظر می‌رسد بررسی این چالش‌ها می‌تواند راه‌گشای مؤثری برای بهینه‌سازی فرایندهای زنجیره‌تأمین به‌عنوان یکی از اساسی‌ترین بخش‌های سازمان‌ها (به‌خصوص سازمان‌های تند مصرف) باشد. شکاف تحقیقاتی مرتبط با موضوع پژوهش در جدول (۱) نشان داده شده است.

جدول (۱): مروری بر برخی از تحقیقات انجام‌شده درخصوص چالش‌های پیاده‌سازی مبتنی بر فناوری‌های دیجیتال

| رویکرد | فناوری اینترنت اشیا | فناوری هوش‌مصنوعی | زنجیره‌تأمین | مطالعه موردی | حوزه | سال | نویسنده |
|---|---|---|---|---|---|---|---|
| ارائه سیستم پشتیبان تصمیم‌گیری | × | × | ✓ | صنعت هتل‌داری | صنایع هوشمند | ۲۰۲۱ | کمر[3] |
| AHP | × | × | ✓ | صنایع تولیدی هند | نسل چهارم صنعتی | ۲۰۱۸ | لوترا و همکاران |
| بررسی مروری | × | ✓ | ✓ | × | کسب و کار هوشمند | ۲۰۱۹ | ویویدی[4] و همکاران |
| بررسی مروری | × | ✓ | ✓ | × | صنایع هوشمند | ۲۰۱۹ | صلاح و همکاران |
| بررسی مروری | × | × | × | × | شهر هوشمند | ۲۰۱۵ | مونزون[5] |
| بررسی مروری | × | × | × | × | شهر هوشمند | ۲۰۲۰ | هسیجا[6] و همکاران |

<div dir="rtl">

ادامه جدول (۱)

| نویسنده | سال | حوزه | مطالعه موردی | زنجیره‌تأمین | فناوری هوش‌مصنوعی | فناوری اینترنت اشیا | رویکرد |
|---|---|---|---|---|---|---|---|
| باکر[1] و همکاران | ۲۰۱۷ | درمان هوشمند | × | × | × | ✓ | ارائه مدل |
| شاه و بات[2] | ۲۰۲۰ | درمان هوشمند | × | × | × | ✓ | ارائه چارچوب مفهومی |
| روب و باهمیا[3] | ۲۰۱۹ | کارخانه هوشمند | صنایع تولیدی | × | × | ✓ | بررسی مروری |
| کیم[4] و همکاران | ۲۰۱۹ | کارخانه هوشمند | × | ✓ | × | ✓ | بررسی مروری |
| نتو[5] و همکاران | ۲۰۲۰ | کارخانه هوشمند | × | × | × | ✓ | تحلیل سلسله مراتبی |
| تحقیق حاضر | ۲۰۲۱ | صنایع هوشمند | صنایع FMCG | ✓ | ✓ | ✓ | تحلیلی سلسله مراتبی غیرخطی فازی |

"زنجیره‌تأمین ۴٫۰ به‌عنوان مجموعه‌ای از فعالیت‌های به‌هم‌پیوسته مربوط به هماهنگی، برنامه‌ریزی و کنترل محصولات و خدمات بین تأمین‌کنندگان و مصرف‌کنندگان" تعریف شده است. هدف آن ایجاد راه‌های جدید برای ایجاد ارزش افزوده برای مشتریان و تأمین‌کنندگان و ایجاد درآمد بیشتر از طریق یکپارچه سازی و هماهنگی فرایندها است (رسول[6] و همکاران،

</div>

---

۲۰۲۱). شش ویژگی یک زنجیره‌تأمین دیجیتال و هوشمند که باید تمامی مشتریان و تأمین‌کنندگانی که زنجیره‌تأمین را تشکیل می‌دهند را پوشش دهد توسط وو[1] و همکاران ارائه شده است. این ویژگی‌ها در جدول (۲) نشان داده شده است.

**جدول(۲): ویژگی‌های زنجیره‌تأمین هوشمند 0/4 (وو و همکاران، ۲۰۱۷)**

| مشخصات | شرح |
|---|---|
| ابزار مند | سیستم‌هایی با حسگرها، برچسب‌های RFID، متر و سایر اجزای یکپارچه که قادر به تولید داده برای تصمیم گیری هستند. |
| به هم پیوسته | اعضای زنجیره‌تأمین، از جمله دارایی‌ها، سیستم‌های فناوری اطلاعات، محصولات و سایر اشیاء هوشمند، کاملاً به هم متصل هستند. |
| هوشمند | سیستم‌های هوشمند قادر به تصمیم گیری به منظور بهینه سازی عملکرد جهانی خود با جمع آوری و تجزیه و تحلیل حجم زیادی از داده‌ها هستند. |
| خودکار | فعالیت‌های خودکار متعددی که با هدف جایگزینی منابع کمتر کارآمد (از جمله نیروی کار) مورد استفاده قرار میگیرند. |
| یکپارچه | فعالیت‌های زنجیره‌تأمین یکپارچه، شامل همکاری بین اعضا، تصمیم گیری با یکدیگر، استفاده از سیستم‌های مشترک و به اشتراک گذاری اطلاعات. |
| خلاقانه | توانایی توسعه و تجمیع ارزش‌های جدید از طریق راه‌حل‌های کارآمدتر. |

این ویژگی‌ها به دلیل فناوری‌های فعال شده توسط نسل چهارم صنعتی است. چهار مورد از این فناوری‌ها اساس فناوری نسل چهارم صنعتی در نظر گرفته می‌شوند:

- **اینترنت اشیا**: هدف اینترنت اشیا گسترش و اتصال اشیاء فیزیکی به اینترنت با ادغام حسگرها، محرک‌ها و سایر دستگاه‌هایی است که داده‌ها را جمع‌آوری، انتقال و پردازش می‌کنند. در یک زنجیره‌تأمین دیجیتال، این فناوری برنامه‌های تجاری و وب (مانند رسانه‌های اجتماعی) را با ماشین‌ها، دستگاه‌ها، محصولات، مواد و افراد کنار هم می‌آورد و ایجاد یک شبکه هوشمند را ممکن می‌سازد که در تمام فرآیندهای کارخانه و مشتریان و تأمین‌کنندگان گسترش می‌یابد (مارول[2]، ۲۰۲۱).

- **سیستم‌های سایبری فیزیکی**: شامل ماشین‌ها، سیستم‌های ذخیره‌سازی و امکانات تولیدی می‌شوند که به‌صورت دیجیتالی توسعه یافته‌اند و یکپارچه‌سازی سرتاسری را بر

---

اساس فناوری‌های اطلاعات و ارتباطات ارائه می‌کنند. آن‌ها با تمرکززدایی، انطباق و رفتار مستقل مشخص می‌شوند و فرصت‌های زنجیره‌تأمین را برای نظارت بر شرایط تولید و فعالیت‌های لجستیکی در زمان واقعی ارائه می‌دهند و امکان پیش‌آگهی و تشخیص و کنترل از راه دور را فراهم می‌کنند (رادانلیو[1] و همکاران، ۲۰۲۱).

● **رایانش ابری**: با هدف ادغام فناوری‌ها یا معماری‌ها برای ارائه یک پلتفرم یا ارائه راه‌حل از طریق اینترنت، امکان دسترسی به آن‌ها را در هر زمان و هر مکان فراهم می‌کند و دید، بینش و انعطاف‌پذیری بی‌سابقه‌ای را نیز ایجاد می‌نماید (صدیق[2] و همکاران، ۲۰۲۱).

● **کلان داده‌ها**: مجموعه بزرگی از داده‌ها را از منابع مختلف، چه سنتی و چه دیجیتال، تشکیل می‌دهند، جایی که می‌توان تحلیل‌ها و اکتشافات مداوم را انجام داد. کلان داده‌ها در پنج بعد – حجم، سرعت، تنوع، صحت و ارزش– متمایز می‌شوند و می‌توانند برای تحلیل‌های توصیفی، پیش‌بینی و تجویزی استفاده شوند. در زنجیره‌تأمین، کلان داده‌ها برنامه‌هایی را ارائه می‌دهند که شامل جریان‌های مواد (مانند وضعیت تولید، نظارت بر فرآیند و کیفیت، مدیریت موجودی، لجستیک، تحقیق و توسعه، و راه‌حل‌های جمعی در عملکردهای تهیه و توزیع)، جریان‌های اطلاعاتی (مانند مدیریت تقاضا، مدیریت رویداد زنجیره‌تأمین، مذاکره فروشنده، مدیریت ریسک، شناسایی مشکل، پشتیبانی تصمیم‌گیری خودکار و مدیریت مشتری) و جریان‌های مالی (مانند تقسیم‌بندی مشتری، مدل‌سازی تقاضا، طراحی مدل کسب‌وکار جدید، قیمت‌گذاری و مجموعه، و جنبه‌های مالی منابع انسانی) می‌شود (نوذری و همکاران، ۲۰۲۱).

فناوری‌های دیگری مانند ساخت افزودنی، اتوماسیون و ربات‌های صنعتی، واقعیت افزوده، امنیت سایبرنتیک، بلاک چین، اینترنت داده‌ها، افراد و خدمات، فناوری‌های معنایی و شبیه‌سازی و مدل‌سازی که بر زنجیره‌تأمین تأثیر می‌گذارند نیز منبع مزیت رقابتی برای زنجیره‌تأمین میباشند و آن را به یک مجموعه هوشمند تبدیل می‌کنند. به لطف این فناوری‌ها و زیرسیستم‌ها و دستگاه‌های آن‌ها، می‌توان کل زنجیره‌تأمین (نه تنها مشتریان و تأمین‌کنندگان، بلکه دارایی‌ها،

---

محصولات و محیط عملیاتی آنها) را یکپارچه کرد و حجم بیشتری از داده‌ها را با کیفیت و سرعت بیشتر تولید کرد. علاوه بر این، فناوری‌ها سازمان‌ها را قادر می‌سازند تا انعطاف‌پذیری، بهره وری، قابلیت اطمینان و پاسخگویی را در عملیات خود افزایش دهند. علاوه بر این، با فعال کردن سازماندهی مجدد کل عملیات در زمان واقعی، سازمان‌ها امکان کاهش اثر شلاقی و هزینه‌های مرتبط با عملیات زنجیره‌تأمین را دارند.

## ۳  مبانی نظری

### ۱-۳ چالش‌های پیاده‌سازی زنجیره‌تأمین ۴٫۰

فناوری به‌عنوان نیروی محرکه در تسریع فرآیندها در مواجهه با انواع فعالیت‌ها، نقش عمده‌ای را در دنیای امروزی ایفا می‌کند، یکی از این شاخه‌های مهم فناوری، فناوری‌های اینترنت اشیا و هوش‌مصنوعی است که امروزه نقش بسزایی در صنایع مختلف ایفا می‌کنند. اما همچنان این فناوری‌ها مانند سایر فناوری‌های نوظهور با چالش‌هایی مواجه هستند که در این مقاله سعی شده است چالش‌های موجود در استفاده از اینترنت اشیا و هوش‌مصنوعی در زنجیره‌تأمین به‌عنوان یکی از مهم‌ترین بخش‌های سازمان‌های تولیدی بررسی شود. اگرچه مفهوم صنایع هوشمند تنها چند سالی است که با حضور فناوری‌هایی مانند اینترنت اشیا و توسعه مفهوم و ابزارهای تجزیه و تحلیل داده‌های بزرگ مطرح شده است و تحقیقاتی در این زمینه انجام شده است، اما پیاده‌سازی یک زنجیره‌تأمین مبتنی بر اینترنت اشیا در نسل چهارم صنعتی چالش‌های زیادی را نیز به‌همراه دارد (دی وس[1] و همکاران، ۲۰۲۱). در این بخش، با استفاده از مرور سیستماتیک ادبیات و مطالعات منتشر شده، مهم‌ترین چالش‌های تأثیرگذار شناسایی و خلاصه شدند.

بدین منظور در گام نخست سؤال پژوهش صورت‌بندی گردید. در این تحقیق سؤالات پژوهش عبارت‌اند از: چالش‌های اصلی تأثیرگذار بر پیاده‌سازی یک زنجیره‌تأمین هوشمند دیجیتال مبتنی بر اینترنت اشیا چه هستند؟ و دارای چه طبقه‌بندی‌ای می‌باشند؟ در همین گام، تعیین جامعه‌ی

---

آماری و بازه‌ی زمانی جستجو نیز تعیین گردید. برای تعیین جامعه‌ی آماری، مقالات نمایه شده در پایگاه‌های نمایه‌سازی اسکوپوس، گوگل اسکولار و آی اس آی وب او ساینس مدنظر قرار گرفت. همچنین با توجه به جدید بودن مفهوم زنجیره‌تأمین نسل ۴ مقالات نمایه شده بین سال‌های ۲۰۱۵ تا ۲۰۲۱ نیز به‌عنوان بازه‌ی زمانی مدنظر قرار گرفت.

درگام دوم مطالعات مرتبط با سؤالات تحقیق شناسایی شدند. برای این منظور باید کلیدواژه مناسب برای جستجو تعیین گردد. کلیدواژه‌های موردنظر برای جستجو، پیاده‌سازی زنجیره‌تأمین دیجیتال، چالش‌های پیاده‌سازی اینترنت اشیا، چالش‌های زنجیره‌تأمین هوشمند، چالش‌های زنجیره‌تأمین مبتنی بر اینترنت اشیا، چالش‌های زنجیره‌تأمین مبتنی بر تحولات دیجیتالی تعیین شدند. به‌منظور استخراج مطالعات مرتبط در این حوزه به پایگاه‌های نمایه‌سازی اسکوپوس، گوگل اسکولار و آی اس آی وب او ساینس مراجعه شد. پس از جستجوی کلیدواژه‌ها در این پایگاه‌ها از ابتدای سال ۲۰۱۵ تا ۲۰۲۱ میلادی[1] تعداد ۲۵۰ عنوان مستند یافت شد.

در گام سوم با تعیین معیارهای شمول، مطالعاتی که باید بررسی گردند شناسایی شدند. معیارهای شمول عبارت‌اند از، زبان انگلیسی، دارابودن یافته‌های کیفی، انتشار در بازه‌ی زمانی ۲۰۱۵ تا ۲۰۲۱ و تأکید بر چالش‌های اصلی تأثیرگذار بر پیاده‌سازی یک زنجیره‌تأمین دیجیتال و هوشمند مبتنی بر اینترنت اشیا؛ بنابراین از تعداد ۲۵۰ مقاله منتشر شده در طی سال‌های بررسی، ۲۰ مورد که به زبان‌هایی غیر از زبان انگلیسی نوشته شده بودند از لیست بررسی حذف شدند. پس از بررسی عنوان مقالات و چکیده‌ی ۲۳۰ مقاله‌ی باقیمانده باتوجه‌به معیارهای شمول و موضوع، سؤال و هدف پژوهش،۱۳۰ مقاله حذف شد و ۱۰۰ مقاله برای بررسی محتوی کامل مقاله در نظر گرفته شد. با بررسی محتوی مقالات، ۶۰ مقاله ویژگی‌های لازم برای استفاده در تحقیق حاضر را نداشته و از روند مطالعه حذف شدند و ۴۰ مقاله دارای ویژگی‌های لازم برای بررسی عمیق‌تر و استفاده در این تحقیق، شناسایی شد.

در گام چهارم کیفیت مقالات مورد ارزشیابی انتقادی قرار گرفت. برای این منظور از روش «برنامه مهارت‌های ارزشیابی حیاتی[1]» استفاده گردید. این ابزار با طرح سؤالاتی بر سه مقوله‌ی استحکام، اعتبار و ارتباط متون تمرکز می‌نماید. برای این منظور با طرح ۱۰ معیار درارتباط با هر مقاله شامل: وضوح اهداف و اهمیت پژوهش، تناسب و تطابق روش پژوهش، تناسب و تطابق طرح پژوهش، تناسب روش انتخاب مشارکت کنندگان، تناسب روش جمع‌آوری داده‌ها، رابطۀ پژوهشگر و مشارکت کنندگان، ملاحظات اخلاقی، دقت تجزیه‌وتحلیل، بیان روشن یافته‌ها و ارزش پژوهش، کیفیت کلی مقاله را ارزیابی می‌نماید. برای این منظور به هر یک از این معیارهای ده‌گانه برای هر مقاله عددی بین ۱ تا ۵ داده می‌شود و در نهایت مقالاتی که در مجموع معیارهای ده‌گانه عددی بالاتر از ۴ دریافت کنند با کیفیت مناسب ارزیابی می‌شوند. در این تحقیق از میان ۴۰ مقاله‌ی بررسی شده توسط شاخص کسپ، کیفیت ۳۲ مقاله بسیار بالا ارزیابی گردید. پس از بررسی مقالات تعداد ۳۰ چالش از متون استخراج گردید که بر پیاده‌سازی زنجیره‌تأمین نسل ۴ اثرگذار می‌باشند.

سپس با استفاده از روش دلفی اصلاح‌شده و با بهره‌گیری از طیف لیکرت ۵ سطحی مطرح شده توسط (آزادی و همکاران، ۲۰۲۰) با نظرخواهی از خبرگان چالش‌های اصلی تأثیرگذار بر پیاده‌سازی زنجیره‌تأمین نسل ۴ استخراج شدند. برای این منظور یک پرسشنامه با پاسخ‌های بسته شامل یک طیف لیکرت ۵ سطحی برای خبرگان این حوزه ارسال گردید. به دلیل دسترسی مؤلفان به شرکت‌های FMCG، چند شرکت فعال در این حوزه به‌عنوان مطالعه موردی در نظر گرفته شدند. خبرگان در میان افراد شاغل با سابقه کاری بالا در حوزه زنجیره‌تأمین و با دانش بالا در زمینه فناوری اطلاعات و آشنا با مفاهیم تحولات دیجیتالی و اینترنت اشیا و هوش‌مصنوعی انتخاب گردیدند. در دور اول دلفی از خبرگان خواسته شد تا نظر خود را در رابطه با میزان اهمیت ۳۰ چالش استخراج شده بر مبنای مقیاس: ۴= خیلی مهم، ۳= مهم، ۲= نسبتا با اهمیت، ۱=کم اهمیت، ۰= بی اهمیت بیان کنند. در پایان دور اول داده‌ها تحلیل شد و برای اجرای دور دوم دلفی مورد

---

استفاده قرار گرفت. چالش‌هایی که موفق به کسب ۷۵ درصد از اجماع جمعی ۴= خیلی مهم و ۳= مهم شدند موردپذیرش قرار گرفتند و سایر مواردی که اجماع جمعی کمتری داشتند برای دور دوم دلفی در نظر گرفته شدند. از میان ۳۰ چالش شناسایی شده، ۷ چالش در دور اول مورد تأیید خبرگان قرار گرفتند. در دور دوم دلفی نیز مانند دور اول عمل شد و در نتیجه ۳ چالش دیگر مورد تأیید خبرگان قرار گرفتند و سایر چالش‌ها از تحقیق حذف شدند. در نتیجه در مجموع ۱۰ چالش اصلی در ۳ دسته بندی شناسایی گردید که در جدول (۳) قابل مشاهده می‌باشند.

**جدول (۳): مهم‌ترین چالش‌های پیاده‌سازی یک زنجیره‌تأمین دیجیتال و هوشمند (زنجیره‌تأمین ۴٫۰)**

| دسته | کد | رفرنس‌ها | چالش |
|---|---|---|---|
| چالش‌های تکنیکی W1 | W11 | (پاپس[1] و همکاران، ۲۰۲۱) (شارما[2] و همکاران، ۲۰۲۱) (نوذری و همکاران، ۲۰۱۸) | پیچیدگی سیستم‌ها |
| | W12 | (مبارک[3] و همکاران، ۲۰۲۱) (نایال[4] و همکاران، ۲۰۲۱) (پنتا[5] و همکاران، ۲۰۲۰) (قد[6] و همکاران، ۲۰۲۰) | چالش‌های تحلیلی و حجم محاسباتی بالا |

**ادامه جدول (۳)**

| دسته | کد | رفرنس‌ها | چالش |
|---|---|---|---|
| | W13 | (لزوچ[7] و همکاران، ۲۰۲۰) (کومار[8] و همکاران، ۲۰۲۱) (بگ[9] و همکاران، ۲۰۲۱) (لی[10] و همکاران، ۲۰۱۹) | امنیت و حریم خصوصی |
| | W14 | (چتری[11] و همکاران، ۲۰۱۸) | چالش‌های اتصال |

---

1 Pappas
2 Sharma
3 Mubarik
4 Nayal
5 Peña
6 Ghode
7 Lezoche
8 Kumar
9 Bag
10 Lee
11 Chhetri

| | | | |
|---|---|---|---|
| | | (داو[1] و همکاران، ۲۰۱۹) | |
| چالش‌های محیطی،مالی،فرهنگی W2 | W21 | (یاداو[2] و همکاران، ۲۰۲۰) (هیدالگو[3] و همکاران،۲۰۱۹) (قج[4] و همکاران، ۲۰۲۰) (دوسو[5]، ۲۰۱۸) | ریسک‌های محیطی |
| | W22 | (سونی و نایک[6]، ۲۰۱۹) (عمر[7] و همکاران، ۲۰۲۱) (گوپتا[8] و همکاران، ۲۰۲۰) | مدیریت انرژی |
| | W23 | (یوانف[9] و همکاران،۲۰۱۸) (که[10] و همکاران، ۲۰۱۹) (مولر[11] و همکاران، ۲۰۱۹) | هزینه سرمایه گذاری |
| | W24 | (که و همکاران، ۲۰۱۹) (مبارک و پتریت[12]،۲۰۲۰) | عدم اعتماد |
| چالش‌های تکنولوژیکی W3 | W31 | (شاو[13] و همکاران،۲۰۲۱) (لیبونی[14] و همکاران، ۲۰۱۹) | عدم وجود دانش و مهارت |
| | W32 | (مکتدیر[15] و همکاران، ۲۰۱۸) (داسیلوا[16] و همکاران، ۲۰۲۰) (رادانلیو[17] و همکاران، ۲۰۱۹) (لوترا[18] و همکاران، ۲۰۲۰) | عدم وجود زیرساخت‌های مناسب |

در ادامه این چالش‌ها را به صورت مختصر شرح داده شده‌اند.

• **پیچیدگی سیستم‌ها:** مفهوم زنجیره‌تأمین هوشمند و دیجیتال در درجه اول فرایندهای زنجیره‌تأمین را به‌عنوان یک سیستم که دارای زیرسیستم‌های متعدد است، بررسی می‌کند این عملکرد زیرسیستم به‌عنوان یک کل در نهایت به آن‌ها اجازه می‌دهد که به شیوه هوشمند و هماهنگ رفتار کند. به عبارتی همچنان که زنجیره‌تأمین یک سیستم پیچیده متشکل از روابط متقابل متنوع و غیرقابل پیش بینی بین زیرسیستم‌های آن است،

---

1 Daú
2 Yadav
3 Hidalgo
4 Ghadge
5 Dossou
6 Sony & Naik
7 Umar
8 Gupta
9 Ivanov
10 Koh
11 Müller
12 Mubarak & Petraite
13 Shao
14 Liboni
15 Moktadir
16 Da Silva
17 Radanliev
18 Luthra

افزوده شدن فناوری‌های دیجیتالی و انواع تکنولوژی‌های متفاوت بر پیچیدگی این سیستم می‌افزاید و بنابراین برقراری ارتباط با این سیستم‌های پیچیده می‌تواند یک چالش اساسی برای زنجیره‌های تأمینی باشد که با استفاده از اینترنت اشیا داده‌ها را دریافت می‌کنند، و تحلیل‌های آن‌ها مبتنی بر هوش‌مصنوعی و یادگیری‌های ماشینی است(پاپس و همکاران، ۲۰۲۱).

- **چالش‌های تحلیلی و حجم محاسباتی بالا**: با پیشرفت تکنولوژی و افزایش روزافزون جریان عظیمی از تولید داده به واسطه حضور فناوری تحول آفیرینی به نام اینترنت اشیا به‌عنوان بزرگترین منابع تولید داده‌های بزرگ و تشکیل پایگاه‌های داده‌های مختلف، تعداد فعالیت‌های محاسباتی در یک زنجره تأمین که خود همواره با چالش‌های محاسباتی بسیاری در بخش‌های تأمین و تولید و فروش دارد، بینهایت افزایش می‌یابد. بنابراین مدیریت کردن این محاسبات و نظارت بر عملکرد صحیح محاسباتی می‌تواند یک چالش مهم در یک زنجیره‌تأمین هوشمند و دیجیتال باشد که باید برای آن تمهیدات معقولی در نظر گرفته شود(موهنتا[1] و همکاران، ۲۰۲۱).

- **امنیت و حریم خصوصی**: از آنجایی‌که ترکیبی از سیستم‌های فیزیکی و دیجیتال در یک کارخانه هوشمند و بر این اساس، در یک زنجیره‌تأمین هوشمند که علاوه بر تولید و نگهداری داده‌ها با استفاده از فناوری اینترنت اشیا، از هوش‌مصنوعی برای تجزیه و تحلیل استفاده می‌کند و امکان همکاری بلادرنگ را برای همه سیستم‌های در حال فعالیت فراهم می‌کند مشغول به کار هستند، لذا همواره خطراتی برای این سیستم‌ها وجود دارد. گسترش سطح حمله با چندین ماشین و دستگاه متصل به یک یا چند شبکه در فرآیندهای هوشمند از جمله این خطرات است. آسیب‌پذیری‌ها در هر یک از این دستگاه‌ها می‌توانند سیستم را برای حمله مجدد و توسط عوامل متفاوت باز کنند. بنابراین شرکت‌ها باید هم آسیب‌پذیری‌های سیستم سازمانی و هم آسیب‌پذیری‌های عملیاتی در سطح ماشین را پیش‌بینی کنند. بسیاری از شرکت‌ها همیشه آمادگی مقابله با این تهدیدات امنیتی را ندارند

---

و این خود از مهمترین چالش‌هایی است که در پیاده‌سازی یک زنجیره‌تأمین دیجیتال و هوشمند باید بدان توجه ویژه ای نمود (فلاح و نوذری، ۲۰۲۰).

• **چالش‌های اتصال**: اتصال سیمی در دنیای صنایع سنتی بسیار محبوب است، ولی با سوق دادن اینترنت اشیا به اتصال بی سیم در ابعاد مختلف صنایع، این امر می‌تواند نشان دهنده تغییر در طراحی زیرساخت شبکه باشد. استفاده از شبکه‌های بی سیم ممکن است به نگرانی‌های امنیتی نیز منجر شود. در اینترنت اشیا، دستگاه‌های دیجیتال از طریق اینترنت به یکدیگر متصل می‌شوند و با یکدیگر ارتباط برقرار می‌کنند و در شبکه‌های کوچک و چند دستگاهی، اتصال‌ها کاملا یکپارچه است. اما زمانی که اینترنت اشیا در مقیاس جهانی مورد استفاده قرار می‌گیرد و تعداد دستگاه‌ها و حسگرها به هم متصل شده و با یکدیگر ارتباط برقرار می‌کنند، مشکلات اتصال به وجود می‌آید. میدانیم که اینترنت فقط یک شبکه نیست، بلکه شامل شبکه‌های ناهمگن با برج‌های سلولی، اتصال آهسته، اتصال سریع، سرورهای پروکسی و فایروال‌ها و شرکت‌های مختلف با استانداردها و فناوری‌های مختلف است که می‌توانند ارتباط را مختل کنند. اتصال به‌عنوان یکی از اجزای مهم اینترنت اشیا در نظر گرفته می‌شود زیرا انتقال داده به یک اتصال خوب بستگی دارد. همچنین تجزیه و تحلیل کلان داده با استفاده از هوش‌مصنوعی و با ارتباطات بسیار، پیچیدگی‌های خاص خود را دارد و یکی از چالش‌های اصلی زنجیره‌تأمین مبتنی بر این فناوری‌ها است(رحیم و همکاران، ۲۰۲۱).

• **ریسک‌های محیطی**: مشاغل به‌طور فزاینده‌ای در برابر خطرات زیست‌محیطی و تغییرات آب و هوایی آسیب‌پذیر هستند. یک زنجیره‌تأمین هوشمند که از اینترنت اشیا و هوش‌مصنوعی پشتیبانی می‌کند باید دارای سیستم‌های مدیریت بلایا بسیار واکنش‌گر و چابک باشد، مانند هشدار دادن به ذینفعان در مورد اقدامات پیشگیری از خطر و کاهش سطح آلودگی. هوش‌مصنوعی و اینترنت اشیا را می‌توان در مرحله طراحی برای به حداقل رساندن خطرات زیست‌محیطی در محیط کسب و کار و همچنین توسعه محصول مورد استفاده قرار داد. پهپادهای خودکار در زمینه‌های مختلفی مانند شناسایی و ثبت خطرات زیست‌محیطی، تنظیم ترافیک برای توزیع محصولات و نظارت بر آلودگی‌های محیطی مورد استفاده قرار گرفته‌اند. سنسورهای کیفیت هوا در پلتفرم‌های آنلاین در فرآیندهای

زنجیره‌تأمین، به ویژه در تولید، می‌توانند از اندازه‌گیری خطرات زیست محیطی پشتیبانی کنند (مونرا[1] و همکاران، ۲۰۲۱).

● **مدیریت انرژی**: مزیت دستگاه‌های اینترنت اشیا این است که اتوماسیون را فعال می‌کنند، با این حال، مقدار قابل توجهی انرژی برای اتصال میلیاردها دستگاه متصل به یکدیگر مورد نیاز است. این نیاز به انرژی، مگر اینکه مدیریت شود، می‌تواند مانعی برای اجرای کامل سیستم‌های اینترنت اشیا باشد. پیشرفت‌های فناوری و تغییرات در عادات مصرف‌کننده منجر به افزایش تقاضای انرژی می‌شود و تولیدکنندگان انرژی اکنون به دنبال کمک هوش‌مصنوعی و اینترنت اشیا برای بهینه‌سازی توزیع انرژی مورد نیاز هستند(ونگ[2] و همکارن، ۲۰۲۱).

● **هزینه سرمایه گذاری**: رکن اصلی پیاده سازییک زنجیره‌تأمین هوشممند مبتنی بر اینترنت اشیا و فناوری هوش‌مصنوعی، تجهیز لوازم و دستگاه‌ها به حسگرها و قطعات سخت افزاری است که داده‌ها را به پلتفرم اینترنت اشیا مخابره می‌نمایند و ابزارهایی که توان محاسباتی بالا دارند. ولی هزینه قطعات موجود نیز بسیار بالا است. از آنجا که برای داشتن یک سیستم یکپارچه هوشمند باید تعداد زیادی از حسگرها را به کار برد، تهیه و نصب آن‌ها بسیار هزینه بر خواهد بود و بنابراین چالش هزینه بر بودن امکانان همواره بر سر راه پیادسازی زنجیره‌های تأمین دیجیتال است(دی وس و همکارن، ۲۰۲۱).

● **عدم اعتماد**: اعتماد اغلب در سطح روابط بین فردی است. در زندگی مدرن هوشمند امروزی، اعتماد بین مردم به‌طور فزاینده ای اعتماد سیستمی است. عدم اعتماد به هوش‌مصنوعی و اینترنت اشیا ممکن است اجرای آن‌ها را در توسعه یک زنجیره‌تأمین قوی و هوشمند کند نماید. اما با مرور ادبیات موضوع می‌توان دریافت که یکی از ویژگی‌های اساسی سیستم‌های مبتنی بر هوش‌مصنوعی، قابلیت اطمینان بالای آنهاست. با این حال، مطالعات نشان می‌دهد که همیشه نگرانی و بی اعتمادی نسبت به سیستم‌های هوشمند مبتنی بر متن در اینترنت اشیا وجود دارد. به ندرت ادعا می‌شود که یک سیستم اینترنت

---

اشیا برای هر محیط، زمینه و هر رویداد غیرمعمولی که سیستم می‌تواند تجربه کند کاملاً دقیق عمل می‌کند. اعتماد به این معنی است که ارزیابی قابلیت اطمینان به شدت به دانش دقیق از زمینه و محیط و انعطاف‌پذیری برای رسیدگی به رویدادها و داده‌های غیر معمول بستگی دارد. به ندرت چنین دانشی وجود خواهد داشت و انعطاف‌پذیری کامل را فراهم می‌کند(الدوواه[1] و همکاران، ۲۰۲۱).

• **عدم وجود دانش و مهارت**: درست است که اینترنت اشیا و فناوری‌های تحول‌آفرین به‌عنوان یک رویکرد بسیار پرطرفدار در دنیا مطرح است اما همچنان این مفهوم در میان بسیاری از مردم، ناآشنا بوده و آگاهی کافی نسبت به آن وجود ندارد. برای مثال مزایایی که اینترنت اشیا در صرفه جویی منابع انرژی به همراه خواهد داشت برای بسیاری از صاحبان کسب و کارها شفاف نشده است. از این رو بدیهی است که تمایل چندانی برای استفاده از این فناوری در بازارهای داخل وجود نداشته و تأمین‌کنندگان این نوع محصولات باید وقت و هزینه زیادی برای یادگیری مصرف کنندگان و جلب توجه آنان بپردازند(چاروال[2]، ۲۰۲۱).

• **عدم وجود زیرساخت‌های مناسب**: یکی دیگر از چالش‌های بسیار مهمی که در رابطه با پیاده‌سازی زنجیره‌تأمین دیجیتال و هوشمند مبتنی بر اینترنت اشیا و فناوری‌های تحول‌آفرین وجود دارد، نیاز به پلتفرم‌های نرم‌افزاری است که دارای قابلیت‌های بالایی در گردآوری داده حسگرها، ذخیره‌سازی، پردازش و تحلیل آن‌ها در بازه‌ زمانی اندک باشند. ابزارهای تحلیلی قدرتمند همواره به زیر ساخت‌های قدرتمند و همین‌طور هزینه کلان نیاز دارند. عدم وجود زیرساخت‌های با قدرت بالا و با توان رقابتی همواره می‌تواند چالش بزرگی برای پیاده‌سازی فرایندهای هوشمند در زنجیره‌تأمین بوده و از جمله دلایل شکست پروژه‌های مبتنی بر فناوری‌های دیجیتال است(هی[3]، ۲۰۲۱).

---

# ۴  روش تحقیق

این پژوهش با رویکردی کمی به دنبال بررسی چالش‌های پیاده‌سازی یک زنجیره‌تأمین دیجیتال هوشمند و مبتنی بر فناوری اینترنت اشیا به منظور کمینه‌سازی چالش‌ها در پیاده‌سازی موفق و مؤثر می‌باشد. در این پژوهش به‌منظور بررسی چالش‌ها ابتدا با استفاده از بررسی ادبیات موضوع و نظرات خبرگان فعال و با استفاده توزیع پرسشنامه و بر مبنای طیف لیکرت ۵ سطحی ۱۰ چالش به‌عنوان مهمترین چالش‌ها استخراج گردیدند. ۲۵ خبره با توجه به دسترسی مؤلفان از میان افراد فعال در شرکت‌های مواد غذایی و دارویی و باتوجه به میزان تجربه و مهارت در خصوص زنجیره‌تأمین و فناوری اطلاعات و تحولات دیجیتالی انتخاب شدند. همچنین ۱۳ تن از اساتید خبره از دانشگاه‌های برتر ایران نیز مورد نظرسنجی قرار گرفتند. به منظور اولویت‌بندی مهمترین چالش‌های پیش رو پیاده‌سازی یک زنجیره‌تأمین ۴٫۰ از یک تکنیک تحلیل سلسله مراتبی فازی استفاده گردید. در این پژوهش از روش آلفای کرونباخ به‌منظور سنجش پایایی پرسشنامه‌های مربوطه و با استفاده از نرم‌افزار spss استفاده شده است. مقدار آلفای کرونباخ کل پرسشنامه‌ها هم در کل ۰٫۹۲۱ بدست آمد که مطلوب ارزیابی می‌شود. با توجه به اینکه در این پژوهش ۱۰ چالش به‌عنوان مهمترین چالش‌ها انتخاب شده‌اند، لذا برای هر کدام از چالش‌ها تحلیل عاملی جداگانه‌ای صورت گرفته است. شکل (۱)، متدلوژی تحقیق را نشان می‌دهد.

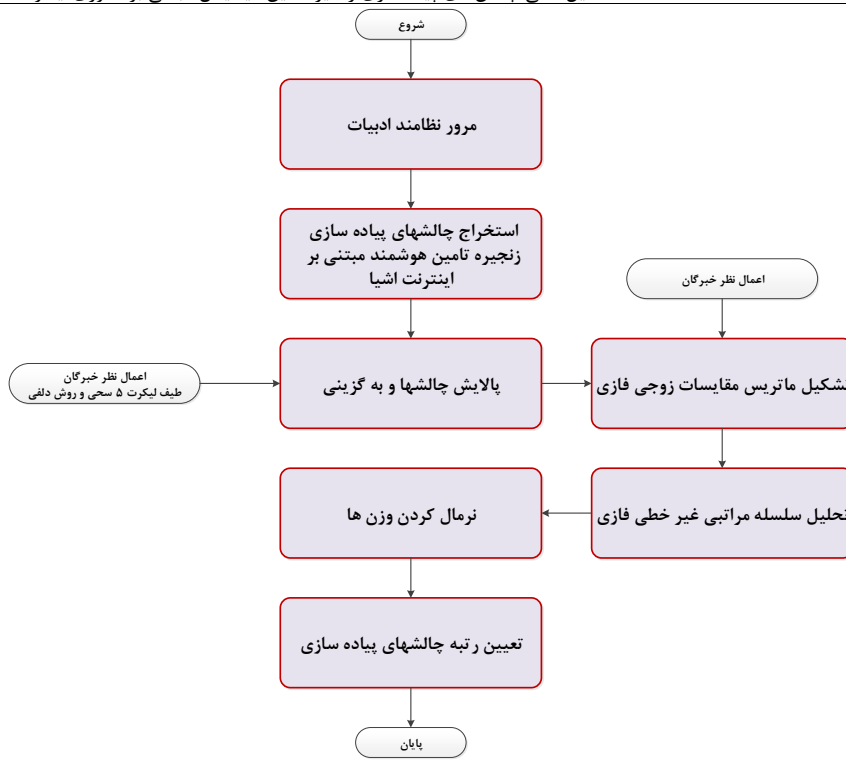

**شکل (۱): متدولوژی تحقیق**

در ادامه روش تحلیل کمی استفاده شده در این پژوهش تشریح شده است.

## ۴-۱ روش غیرخطی رتبه‌بندی فازی میخایلوف

در این پژوهش از یک روش تحلیل سلسله مراتبی غیر خطی فازی که به روش میخایلوف معرف است (فلاح و نوذری، ۲۰۲۰)، به‌منظور اولویت‌بندی مهمترین چالش‌های پیاده‌سازی یک زنجیره‌تأمین دیجیتال مبتنی بر فناوری‌های دیجیتال استفاده شده است. مراحل استفاده از این روش به شرح زیر است:

۱-  ترسیم ساختار سلسله مراتبی: که در جدول (۱) نشان داده شده است.

۲-  تشکیل ماتریس مقایسات زوجی فازی: ماتریس‌های توافقی قضاوت فازی بر اساس نظرات خبرگان تشکیل می‌گردد. به همین دلیل از اعداد فازی در بیان ترجیحات خبرگان

در این پژوهش استفاده شده است. متغیرهای زبانی و مقیاس فازی آن‌ها در جدول (۴) ارائه شده است.

**جدول (۴): متغییرهای زبانی برای مقایسات زوجی**

| مقیاس فازی مثلثی | متغیر زبانی |
|---|---|
| (1,2,3) | خیلی کم |
| (2,3,4) | کم |
| (3,4,5) | متوسط |
| (4,5,6) | زیاد |
| (5,6,7) | خیلی زیاد |

همانگونه که در جدول بالا مشاهده می‌شود، برای مقیاس‌های زبانی از اعداد مثلثی فازی استفاده شده است که یکی از انواع اعداد فازی است که با سه عدد حقیقی به صورت $F = (l, m, u)$ نمایش داده می‌شود. کران بالا که با $u$ نشان داده می‌شود بیشینه مقادیری است که عدد فازی $F$ می‌تواند اختیار کند. کران پایین که با $l$ نشان داده می‌شود کمینه مقادیری است که عدد فازی $F$ می‌تواند اختیار کند. مقدار $m$ محتمل‌ترین مقدار یک عدد فازی است.

۳- تدوین مدل و حل مسئله : در این روش فرض می‌شود مقایسه‌های زوجی فازی به صورت اعداد فازی مثلثی است. بردار قطعی وزن $w = (w_1, w_2, \ldots, w_n)$ به گونه‌ای استخراج می‌شودکه نرخ اولویت تقریباً در محدوده قضاوت‌های فازی ابتدایی قرار گیرد. به عبارت دیگر وزن‌ها طوری تعیین می‌شود که رابطه زیر برقرار باشد.

$$l_{ij} \leq \frac{w_i}{w_j} \leq u_{ij} \qquad (1)$$

هر بردار وزنی قطعی $(w)$ با درجه‌ای در نامعادلات فازی فوق صدق می‌کند که از طریق تابع عضویت خطی رابطه زیر قابل اندازه‌گیری است:

$$\mu_{ij}\left(\frac{w_i}{w_j}\right) = \begin{cases} \dfrac{(w_i / w_j) - l_{ij}}{m_{ij} - l_{ij}} & \dfrac{w_i}{w_j} \le m_{ij} \\[2mm] \dfrac{u_{ij} - (w_i / w_j)}{u_{ij} - m_{ij}} & \dfrac{w_i}{w_j} \le m_{ij} \end{cases} \qquad (2)$$

با در نظر گرفتن شکل خاص توابع عضویت، مسئله اولویت‌بندی فازی به یک مساله بهینه‌سازی غیرخطی به شکل زیر تبدیل می‌شود.

$$\begin{aligned} &\max\ \lambda \\ &Subject\ to: \\ &(m_{ij} - l_{ij})\lambda w_j - w_i + l_{ij} w_j \le 0 \\ &(u_{ij} - m_{ij})\lambda w_j + w_i - u_{ij} w_j \le 0 \qquad (3) \\ &i = 1,2,...,n-1\ ,\ \ j = 2,3,...,n\ ,\ \ j > i, \\ &\sum_{k=1}^{n} w_k = 1 \qquad w_k > 0, \qquad k = 1,2,...,n \end{aligned}$$

با توجه به غیر خطی بودن رابطه ۳، بدیهی است که حل آن بدون استفاده از نرم افزار امکان‌پذیر نیست. لذا برای حل مدل‌های ایجاد شده در این پژوهش از نرم افزار GAMS استفاده شده است. مقادیر بهینه مثبت برای شاخص $\lambda$ (تابع هدف) نشان دهنده این است که تمام نسبت‌های وزن‌ها کاملا در قضاوت اولیه صدق می‌کنند، اما در صورت منفی‌بودن این شاخص، می‌توان فهمید که قضاوت‌های فازی قویا ناسازگار بوده و نسبت‌های وزنی تقریبا در این قضاوت‌ها صدق کرده است.

## ۵  یافته‌های تحقیق

مراحل مربوط به ارزیابی و رتبه‌بندی چالش‌های پیاده‌سازی زنجیره‌تأمین مبتنی اینترنت اشیا و فناوری‌های تحول‌آفرین در این مطالعه به دو بخش اصلی تقسیم می‌شود:

۱- تعیین ماتریس مقایسه‌های زوجی بر اساس ادغام نظرات خبرگان

۲- استفاده از مدل سازی ریاضی به منظور رتبه بندی و به دست آوردن وزن چالش‌ها در مدل تحقیق.

به‌منظور ارزیابی و اولویت‌بندی چالش‌های موجود در این پژوهش، پرسش‌نامه‌های فازی با استفاده از متغیرهای زبان برای ۲۵ نفر از کارشناسان فعال در صنعت و همچنین ۱۳ نفر از کارشناسان دانشگاهی ارسال شد. ۲۸ پرسش‌نامه تکمیل و دریافت شد. این جداول مقایسات زوجی در جداول ۵ تا ۸ نشان داده شده است. از این جداول برای محاسبات به روش میخائیلوف استفاده شده‌است.

جدول (۵): ماتریس مقایسات زوجی برای دسته بندی کلی

|  | W1 | | | W2 | | | W3 | | |
|---|---|---|---|---|---|---|---|---|---|
| W1 | - | - | - | - | - | - | - | - | - |
| W2 | 2.1 | 2.7 | 3.8 | - | - | - | - | - | - |
| W3 | 1.5 | 1.75 | 2.5 | 3.1 | 3.95 | 5.12 | - | - | - |

جدول (۶): ماتریس مقایسات زوجی برای چالش‌های تکنیکی

|  | W11 | | | W12 | | | W13 | | | W14 | | |
|---|---|---|---|---|---|---|---|---|---|---|---|---|
| W11 | - | - | - | - | - | - | - | - | - | - | - | - |
| W12 | 3.1 | 4.2 | 5.1 | - | - | - | - | - | - | - | - | - |
| W13 | 2.1 | 2.8 | 4.7 | 2.3 | 3.1 | 4.2 | - | - | - | - | - | - |
| W14 | 3.1 | 3.5 | 5.4 | 3.1 | 3.5 | 4.5 | 2.1 | 2.45 | 3.21 | - | - | - |

جدول (۷): ماتریس مقایسات زوجی برای چالش‌های محیطی مالی و فرهنگی

|  | W21 | | | W22 | | | W23 | | | W24 | | |
|---|---|---|---|---|---|---|---|---|---|---|---|---|
| W21 | - | - | - | - | - | - | - | - | - | - | - | - |
| W22 | 2.5 | 3.5 | 4.2 | - | - | - | - | - | - | - | - | - |
| W23 | 2.8 | 3.1 | 3.9 | 2.25 | 3.4 | 4.9 | - | - | - | - | - | - |
| W24 | 3.1 | 3.25 | 3.9 | 2.35 | 3.41 | 4.25 | 1.25 | 2.47 | 4.31 | - | - | - |

جدول (۸): ماتریس مقایسات زوجی برای چالش‌های تکنولوژیکی

|  | W31 | | | W32 | | |
|---|---|---|---|---|---|---|
| W31 | - | - | - | - | - | - |
| W32 | 2.5 | 3.47 | 4.25 | - | - | - |

با قرار دادن داده‌های حاصل از جداول ۵ تا ۸ در مدل غیر خطی(۳) به‌عنوان مدل ارائه دهنده وزن‌ها و رتبه‌ها بر اساس تحلیل سلسله مراتبی و حدل مدل با استفاده از نرم افزار GAMS، می‌توان وزن و رتبه هر کدام از چالش‌ها را در ابعاد کلی و همین‌طور در دسته‌های انحصاری به

دست آورد. نتایج محاسباتی مرتبط با حل مدل غیر خطب برای دسته‌های کلی و تک‌تک چالش‌ها در جداول ۹ تا ۱۲ نشان داده شده است.

### جدول (۹): وزن و رتبه‌بندی دسته‌های اصلی

| تابع هدف( $\lambda$ ) | رتبه | وزن | کد | دسته |
|---|---|---|---|---|
| | ۱ | ۰.۳۷۳۸۸۷ | W1 | تکنیکی |
| ۰.۴۳۷۴ | ۳ | ۰.۲۸۱۲۱۰ | W2 | محیطی، مالی و فرهنگی |
| | ۲ | ۰.۳۴۷۱۱۱ | W3 | تکنولوژیکی |

### جدول (۱۰): وزن و رتبه‌بندی چالش‌های تکنیکی

| تابع هدف( $\lambda$ ) | رتبه | وزن | کد | چالش |
|---|---|---|---|---|
| | ۲ | ۰.۲۵۸۸۱۱ | W11 | پیچیدگی سیستم‌ها |
| ۰.۳۲۱۴ | ۴ | ۰.۱۶۵۹۹۸ | W12 | چالش‌های تحلیلی و حجم محاسباتی بالا |
| | ۱ | ۰.۳۸۷۸۴۹ | W13 | امنیت و حریم خصوصی |
| | ۳ | ۰.۱۹۴۳۰۶ | W14 | چالش‌های اتصال |

### جدول (۱۱): وزن و رتبه بندی چالش‌های محیطی، مالی و فرهنگی

| تابع هدف( $\lambda$ ) | رتبه | وزن | کد | چالش |
|---|---|---|---|---|
| | ۱ | ۰.۲۷۱۱۹۴ | W21 | ریسک‌های محیطی |
| ۰.۲۵۴۱ | ۴ | ۰.۲۰۹۳۸۰ | W22 | مدیریت انرژی |
| | ۲ | ۰.۲۶۸۷۷۷ | W23 | هزینه سرمایه گذاری |
| | ۳ | ۰.۲۵۶۸۱۴ | W24 | عدم اعتماد |

### جدول (۱۲): وزن و رتبه‌بندی چالش‌های تکنولوژیکی

| تابع هدف( $\lambda$ ) | رتبه | وزن | کد | دسته |
|---|---|---|---|---|
| | ۲ | ۰.۳۶۳۷۷۵ | W31 | عدم وجود دانش و مهارت |
| ۰.۴۲۵۱ | ۱ | ۰.۶۳۶۲۲۵ | W32 | عدم وجود زیرساخت‌های مناسب |

همان‌گونه که در جداول ۹ تا ۱۲ مشاهده می‌شود مقدار مثبت برای شاخص سازگاری بیانگر سازگاری قابل قبول ماتریس‌ها است.بعد از بدست آورن وزن دسته‌های کلی و وزن چالش‌ها در دسته‌های خاص، می‌توانیم با نرمال کردن وزن‌ها، وزن کلی همه چالش‌ها را بدون در نظر گرفتن دسته و همین‌طور رتبه کلی آن‌ها را بدست آوریم. نتایج محاسباتی نرمال شده در جدول (۱۳) نشان داده شده است.

جدول (۱۳): وزن نرمال و رتبه چالش‌های پیاده‌سازی زنجیره‌تأمین دیجیتال مبتنی بر اینترنت اشیا

| رتبه | وزن نرمال شده | وزن | چالش | وزن | دسته |
|---|---|---|---|---|---|
| 4 | 0.096766 | 0.258811 | پیچیدگی سیستم‌ها | | |
| 9 | 0.062065 | 0.165998 | چالش‌های تحلیلی و حجم محاسباتی بالا | 0.373887 | تکنیکی |
| 2 | 0.145012 | 0.387849 | امنیت و حریم خصوصی | | |
| 7 | 0.072648 | 0.194306 | چالش‌های اتصال | | |
| 5 | 0.076263 | 0.271194 | ریسک‌های محیطی | | |
| 10 | 0.05888 | 0.209380 | مدیریت انرژی | 0.28121 | محیطی، مالی و فرهنگی |
| 6 | 0.075583 | 0.268777 | هزینه سرمایه گذاری | | |
| 8 | 0.072219 | 0.256814 | عدم اعتماد | | |
| 3 | 0.12627 | 0.363775 | عدم وجود دانش و مهارت | 0.347111 | تکنولوژیکی |
| 1 | 0.220841 | 0.636225 | عدم وجود زیرساخت‌های مناسب | | |

همان‌گونه که در جدول (۱۳) مشاهده می‌شود، چالش عدم وجود زیرساخت‌های مناسب و همین‌طور چالش‌های امنیتی از مهم‌ترین چالش‌های پیاده‌سازی زنجیره‌تأمین ۴٫۰ در عصر فناوری‌های دیجیتالی می‌باشند.

# ۶ نتیجه‌گیری و پیشنهادات

زنجیره‌تأمین هوشمند دیجیتال طیفی از فناوری‌ها را معرفی می‌کند که به سازمان‌ها اجازه می‌دهد عملکرد فعالیت‌های داخلی و خارجی خود را افزایش دهند و در نتیجه مزیت رقابتی به دست آورند. بنابراین زنجیره‌تأمین قدرت گرفته با فناوری‌هایی مانند اینترنت اشیا و هوش‌مصنوعی می‌توان تسهیلگر بزرگی برای سازمان‌ها از مرحله تأمین تا فروش باشد. زنجیره‌تأمین مدرن روز به روز پیچیده‌تر می‌شود. اما ریسک‌های پنهان همچنان تهدیدی مهم برای صنعت محسوب می‌شوند. حتی با وجود تمام فناوری‌های جدید که به صنعت راه پیدا کرده‌اند، کسب‌وکارها باید از این خطرات پنهان آگاه باشند و بدانند که چگونه واکنش مناسب نشان دهند. کسب‌وکارها از هر نوع باید قابلیت مشاهده زنجیره‌تأمین، خطرات سایبری و بلایای طبیعی را همیشه در ذهن داشته باشند. همه این عوامل یا حتی یکی از آن‌ها می‌تواند تأثیر قابل توجهی بر سود شرکت داشته باشد. پیشرفت فناوری راه خود را به زنجیره‌تأمین باز

می‌کند و کسب و کارها را مجبور می‌کند که دائماً سیستم‌ها را تغییر دهند و این امر مدیران زنجیره‌تأمین را ملزم می‌کنند که دائماً استعدادهای خود را تقویت کنند و با این نوع اختلالات دیجیتال سازگار شوند. اگرچه هماهنگی فعالیت‌ها، سیستم‌ها و سایر منابع داخلی یک شرکت با مشتریان و تأمین‌کنندگان امکان کسب منافع در عملکرد، کیفیت و هزینه‌ها را فراهم می‌آورد، اما می‌تواند چالش‌های بزرگی را برای شرکت‌ها به همراه داشته باشد.درک این چالش‌ها و مدیریت کردن آن‌ها می‌تواند در پیاده‌سازی مؤثرتر آن‌ها کمک کننده باشد.زیرا با کاهش میزان چالش‌های پیاده‌سازی و استفاده از این تحولات دیجیتالی محصولات در زنجیره‌تأمین شرکت‌ها به‌عنوان یکی از مهم‌ترین بخش‌های سازمان‌های تولیدی می‌توان بیشترین استفاده را از مزایای دیجیتالی شدن نمود. لذا در این پژوهش کوشیده شد تا مهم‌ترین چالش‌های پیش روی پیاده‌سازی یک زنجیره‌تأمین دیجیتال و هوشمند مبتنی بر فناوری اینترنت اشیا مورد بررسی قرار گیرد. در این پژوهش ابتدا با استفاده از بررسی مرور ادبیات چالش‌های پیاده‌سازی استخراج گردید و سپس با استفاده از نظرات خبرگان فعال در صنعت و همین‌طور نظرات برخی از اساتید دانشگاهی مهم‌ترین چالش‌ها مشخص گردید. در این مطالعه برخی از شرکت‌های تند مصرف به‌دلیل دسترسی مؤلفان به این شرکت‌ها و با توجه به ویژگی‌های اساسی این صنایع و اهمیت بالای محصولات آن‌ها در زندگی روز مره انسان‌ها به‌عنوان مطالعه موردی در نظر گرفته شدند. به منظور تعیین رتبه این چالش‌ها برای درک اهمیت آن‌ها از یک روش تحلیل سلسله مراتبی غیر خطی فازی استفاده شد. نتایج نشان داد که چالش‌های مرتبط به نبود زیرساخت‌های مناسب از مهم‌ترین چالش‌هایی است که پیاده‌سازی یک سیستم مبتنی بر تحولات دیجیتالی بدان نیاز دارد. این بخش از جمله الزامات اساسی پیاده‌سازی در هر فناوری دیگری نیز می‌تواند باشد طوری که بسیاری از شرکت‌ها بدون وجود زیرساخت‌های (تکنولوژیکی و فنی و همین‌طور عدم وجود آمادگی لازم در سازمان برای پذیرش فناوری) مناسب در مسیر پیاده‌سازی گام بر نمی دارند. این زیرساخت‌ها یکی از مؤلفه‌های بلوغ سازمانی می‌باشد که خود می‌تواند تأییدکننده رتبه بالای این گزینه باشد. چالش‌های امنیتی در رتبه دوم قرار گرفته که با توجه به حضور فناوری اینترنت و شبکه می‌توان دریافت که مباحث مرتبط با امنیت سایبری چه اهمیت بالایی می‌تواند داشته باشد. امنیت سایبری و خطرات مربوط به حفظ حریم خصوصی مهم‌ترین نگرانی محققان و متخصصان امنیتی است.این امر چالش‌های قابل توجهی برای بسیاری از سازمان‌های

تجاری همواره به همراه دارد. حملات رایج شایع امنیت سایبری آسیب‌پذیری فناوری‌های اینترنت اشیا را نشان داده است. این آسیب‌پذیری صرفاً به این دلیل است که ارتباط متقابل شبکه‌ها در اینترنت اشیا امکان دسترسی از طریق اینترنت ناشناس و غیرقابل اعتماد را فراهم می‌کند که نیاز به راه‌حل‌های جدید امنیتی دارد. از بین همه چالش‌هایی که شناخته شده است، هیچ یک از آن‌ها تأثیر قابل توجهی در سازگاری اینترنت اشیا، مانند امنیت و حریم خصوصی ندارند. با این حال، جای تأسف است که کاربران اغلب تأیید لازم درباره تأثیرات امنیتی را تا زمان وقوع مشکل ندارند، و باعث خسارات زیادی مانند از دست دادن اطلاعات مهم می‌شوند. در مرحله بعدی عدم وجود دانش کافی از مهم‌ترین چالش‌هایی در پیاده‌سازی زنجیره‌تأمین دیجیتال مبتنی بر اینترنت اشیا باید بدان‌ها توجه کرد. بنابراین به نظر می‌رسد که باید با افزایش دانش تخصصی در دانشگاه علاوه بر اینکه باید با تربیت نیروی متخصص در جهت پیشبرد اهداف هوشمندسازی سازمان گام برداشت، باید دانش عمومی در خصوص تحولات دیجیتالی را در عموم کارکنان سازمان‌ها نیز رشد داد تا آمادگی سازمان برای پذیرش فناوری‌ها افزایش یافته و اعتماد نسبت به استفاده از این فناوری‌ها در چرخه‌های صنعتی نیز افزایش یابد. بنابراین می‌توان دریافت که مدیریت کردن این چالش‌ها از مهم‌ترین وظایف مدیران توسعه فناوری در سازمان‌هاست و می‌تواند چابکی و پایداری زنجیره‌تأمین را در حد قابل قبولی افزایش دهد.

لازم به ذکر است که از آنجایی‌که بررسی این مقاله علاوه بر وجود و پیاده‌سازی زیرساخت‌های تکنولوژیکی در صنایع مختلف نیاز به حضورمنابع انسانی توانمند و خبره برای درک بهتر مفاهیم مرتبط با تأثیرات تحولات دیجیتالی دارد، لذا رفع این محدودیت‌ها و آموزش نیروهای خبره و آشنا به تکنولوژی‌های روز دنیا می‌تواند بر دقت پژوهش بیفزاید. همچنین حضور تکنولوژی اینترنت اشیا در زنجیره‌تأمین سایر صنایع به‌عنوان مطالعه موردی نیز می‌تواند مدل وچارچوب قدرتمندتری را ارائه دهد. بنابراین به‌عنوان پیشنهاد می‌توان در پژوهش‌های آتی علاوه بر بررسی حضور فناوری اینترنت اشیا در زنجیره‌تأمین سایر صنایع (برای ارائه اولویت‌بندی قدرتمندتر و دقیقتر)، چالش‌های حضور بسیاری از فناوری‌های تحول آفرین مانند زنجیره بلوکی، یادگیری ماشین و چالش‌های استفاده ترکیبی از این فناوری‌ها (مانند ترکیب هوش‌مصنوعی و اینترنت اشیا که به‌عنوان هوش‌مصنوعی چیزها شناخته می‌شود) را مورد بررسی و واکاوی قرار دارد.

# ۷ مراجع

Abou Kamar, M. (2021). Transforming Hotel Supply Chain Using Intelligent Decision Support System: Prospects and Challenges. *Journal of Association of Arab Universities for Tourism and Hospitality*, *20*(2), 216-246.

Aldowah, H., Rehman, S. U., & Umar, I. (2021). Trust in IoT Systems: A Vision on the Current Issues, Challenges, and Recommended Solutions. *Advances on Smart and Soft Computing*, 329-339.

Azadi, T., Sadoughi, F., & Khorasani-Zavareh, D. (2020). Using modified Delphi method to propose and validate the components of a child injury surveillance system for Iran. *Chinese Journal of Traumatology*, *23*(5), 274-279.

Bag, S., Telukdarie, A., Pretorius, J. H. C., & Gupta, S. (2018). Industry 0,4 and supply chain sustainability: framework and future research directions. *Benchmarking: An International Journal*.

Baker, S. B., Xiang, W., & Atkinson, I. (2017). Internet of things for smart healthcare: Technologies, challenges, and opportunities. *IEEE Access*, *5*, 26521-26544.

Banerjee, A., Lücker, F., & Ries, J. M. (2021). An empirical analysis of suppliers' trade-off behaviour in adopting digital supply chain financing solutions. *International Journal of Operations & Production Management*.

Chawla, S. (2021). Design and implementation of IoT based Low cost, effective learning mechanism for empowering STEM education in India. *Turkish Journal of Computer and Mathematics Education (TURCOMAT)*, *12*(13), 125-133.

Chen, B., Wan, J., Shu, L., Li, P., Mukherjee, M., & Yin, B. (2017). Smart factory of industry 0,4: Key technologies, application case, and challenges. *Ieee Access*, *6*, 6505-6519.

Chhetri, S. R., Faezi, S., Rashid, N., & Al Faruque, M. A. (2018). Manufacturing supply chain and product lifecycle security in the era of industry 0,4. *Journal of Hardware and Systems Security*, *2*(1), 51-68.

Da Silva, V. L., Kovaleski, J. L., Pagani, R. N., Silva, J. D. M., & Corsi, A. (2020). Implementation of Industry 0/4 concept in companies: Empirical evidences. *International Journal of Computer Integrated Manufacturing*, *33*(4), 325-342.

Daú, G., Scavarda, A., Scavarda, L. F., & Portugal, V. J. T. (2019). The healthcare sustainable supply chain 0/4: The circular economy transition conceptual framework with the corporate social responsibility mirror. *Sustainability*, *11*(12), 3259.

de Vass, T., Shee, H., & Miah, S. (2021). IoT in Supply Chain Management: Opportunities and Challenges for Businesses in Early Industry 0/4 Context. *Operations and Supply Chain Management: An International Journal*, *14*(2), 148-161.

Dossou, P. E. (2018). Impact of Sustainability on the supply chain 0,4 performance. *Procedia Manufacturing*, *17*, 452-459.

Dwivedi, Y. K., Hughes, L., Ismagilova, E., Aarts, G., Coombs, C., Crick, T., ... & Williams, M. D. (2019). Artificial Intelligence (AI): Multidisciplinary perspectives on emerging challenges, opportunities, and agenda for research, practice and policy. *International Journal of Information Management*, 101994.

ESSAAIDI, P. M. (2021). *Developing a Smart and Integrated Supply Chain for Hospitals* (Doctoral dissertation, Université Mohammed V de Rabat).

Fallah, M., & Nozari, H. (2020). Quantitative Analysis of Cyber Risks in IoT-Based Supply Chain (FMCG Industries). *Journal of Decisions & Operations Research*, *5*(4).

Ghadge, A., Kara, M. E., Moradlou, H., & Goswami, M. (2020). The impact of Industry 0.4 implementation on supply chains. *Journal of Manufacturing Technology Management*.

Ghode, D., Yadav, V., Jain, R., & Soni, G. (2020). Adoption of blockchain in supply chain: an analysis of influencing factors. *Journal of Enterprise Information Management*.

Hasija, S., Shen, Z. J. M., & Teo, C. P. (2020). Smart city operations: Modeling challenges and opportunities. *Manufacturing & Service Operations Management*, *22*(1), 203-213.

He, W. (2021). IoT System for Intelligent Firefighting in the Electric Power Industry. *Journal of Shanghai Jiaotong University (Science)*, *26*(5), 686-689.

Ivanov, D., Sethi, S., Dolgui, A., & Sokolov, B. (2018). A survey on control theory applications to operational systems, supply chain management, and Industry 0.4. *Annual Reviews in Control*, *46*, 134-147.

Kim, M., Lee, J., & Jeong, J. (2019, July). Open source based industrial iot platforms for smart factory: Concept, comparison and challenges. In *International Conference on Computational Science and Its Applications* (pp. 105-120). Springer, Cham.

Koh, L., Orzes, G., & Jia, F. J. (2019). The fourth industrial revolution (Industry 0.4): technologies disruption on operations and supply chain management. *International Journal of Operations & Production Management*.

Kumar, A., Choudhary, S., Garza-Reyes, J. A., Kumar, V., Rehman Khan, S. A., & Mishra, N. (2021). Analysis of critical success factors for implementing industry 0.4 integrated circular supply chain–Moving towards sustainable operations. *Production Planning & Control*, 1-15.

Lee, J., Azamfar, M., & Singh, J. (2019). A blockchain enabled Cyber-Physical System architecture for Industry 0.4 manufacturing systems. *Manufacturing letters*, *20*, 34-39.

Lezoche, M., Hernandez, J. E., Díaz, M. D. M. E. A., Panetto, H., & Kacprzyk, J. (2020). Agri-food 0.4: A survey of the supply chains and technologies for the future agriculture. *Computers in Industry*, *117*, 103187.

Liboni, L. B., Cezarino, L. O., Jabbour, C. J. C., Oliveira, B. G., & Stefanelli, N. O. (2019). Smart industry and the pathways to HRM 0.4: implications for SCM. *Supply Chain Management: An International Journal*.

Lim, C., Kim, K. J., & Maglio, P. P. (2018). Smart cities with big data: Reference models, challenges, and considerations. *Cities*, *82*, 86-99.

Luthra, S., Kumar, A., Zavadskas, E. K., Mangla, S. K., & Garza-Reyes, J. A. (2020). Industry 0,4 as an enabler of sustainability diffusion in supply chain: an analysis of influential strength of drivers in an emerging economy. *International Journal of Production Research*, *58*(5), 1505-1521.

Marwedel, P. (2021). *Embedded system design: embedded systems foundations of cyber-physical systems, and the internet of things* (p. 433). Springer Nature.

Mohanta, B. K., Satapathy, U., & Jena, D. (2021). Addressing Security and Computation Challenges in IoT Using Machine Learning. In *Advances in Distributed Computing and Machine Learning* (pp. 67-74). Springer, Singapore.

Moktadir, M. A., Ali, S. M., Kusi-Sarpong, S., & Shaikh, M. A. A. (2018). Assessing challenges for implementing Industry 0,4: Implications for process safety and environmental protection. *Process Safety and Environmental Protection*, *117*, 730-741.

Monzon, A. (2015, May). Smart cities concept and challenges: Bases for the assessment of smart city projects. In *2015 international conference on smart cities and green ICT systems (SMARTGREENS)* (pp. 1-11). IEEE.

Mubarak, M. F., & Petraite, M. (2020). Industry 0,4 technologies, digital trust and technological orientation: What matters in open innovation?. *Technological Forecasting and Social Change*, *161*, 120332.

Mubarik, M. S., Naghavi, N., Mubarik, M., Kusi-Sarpong, S., Khan, S. A., Zaman, S. I., & Kazmi, S. H. A. (2021). Resilience and cleaner production in industry 0,4: Role of supply chain mapping and visibility. *Journal of Cleaner Production*, *292*, 126058.

Múnera, D., Aguirre, J., & Gómez, N. G. (2021). IoT-based air quality monitoring systems for smart cities: A systematic mapping study. *International Journal of Electrical & Computer Engineering (2088-8708)*, *11*(4).

Nahr, J. G., Nozari, H., & Sadeghi, M. E. (2021). Green supply chain based on artificial intelligence of things (AIoT). *International Journal of Innovation in Management, Economics and Social Sciences*, *1*(2), 56-63.

Nayal, K., Raut, R. D., Queiroz, M. M., Yadav, V. S., & Narkhede, B. E. (2021). Are artificial intelligence and machine learning suitable to tackle the COVID-19 impacts? An agriculture supply chain perspective. *The International Journal of Logistics Management*.

Neto, A. A., da Silva, E. R., Souza, A., Deschamps, F., de Lima, E. P., & da Costa, S. E. G. (2020). Fault prediction as a service in the smart factory: addressing common challenges for an effective implementation. *IFAC-PapersOnLine*, *53*(2), 10743-10748.

Nozari, H., & Szmelter, A. (Eds.). (2018). *Global supply chains in the pharmaceutical industry*. IGI Global.

Nozari, H., Fallah, M., Kazemipoor, H., & Najafi, S. E. (2021). Big data analysis of IoT-based supply chain management considering FMCG industries. *Business Informatics*, *15*(1), 78-96.

Nozari, H., Najafi, E., Fallah, M., & Hosseinzadeh Lotfi, F. (2019). Quantitative analysis of key performance indicators of green supply chain in FMCG industries using non-linear fuzzy method. *Mathematics*, *7*(11), 1020.

Pappas, N., Caputo, A., Pellegrini, M. M., Marzi, G., & Michopoulou, E. (2021). The complexity of decision-making processes and IoT adoption in accommodation SMEs. *Journal of Business Research*, *131*, 573-583.

Pîrvu, B. C., & Zamfirescu, C. B. (2017, August). Smart factory in the context of 4th industrial revolution: challenges and opportunities for Romania. In *IOP conference series: Materials science and engineering* (Vol. 227, No. 1, p. 012094). IOP Publishing.

Radanliev, P., De Roure, D., Van Kleek, M., Santos, O., & Ani, U. (2021). Artificial intelligence in cyber physical systems. *AI & society*, *36*(3), 783-796.

Radanliev, P., Roure, D. C. D., Nurse, J., Montalvo, R. M., & Burnap, P. (2019). Supply Chain Design for the Industrial Internet of Things and the Industry 0.4.

Rahim, M. A., Rahman, M. A., Rahman, M. M., Asyhari, A. T., Bhuiyan, M. Z. A., & Ramasamy, D. (2021). Evolution of IoT-enabled connectivity and applications in automotive industry: A review. *Vehicular Communications*, *27*, 100285.

Ramirez-Peña, M., Abad Fraga, F. J., Salguero, J., & Batista, M. (2020). Assessing sustainability in the shipbuilding supply chain 0.4: A systematic review. *Sustainability*, *12*(16), 6373.

Ramirez-Peña, M., Abad Fraga, F. J., Salguero, J., & Batista, M. (2020). Assessing sustainability in the shipbuilding supply chain 0.4: A systematic review. *Sustainability*, *12*(16), 6373.

Rasool, F., Greco, M., & Grimaldi, M. (2021). Digital supply chain performance metrics: a literature review. *Measuring Business Excellence*.

Rüb, J., & Bahemia, H. (2019, June). A review of the literature on smart factory implementation. In *2019 IEEE International Conference on Engineering, Technology and Innovation (ICE/ITMC)* (pp. 1-9). IEEE.

Sadeeq, M. M., Abdulkareem, N. M., Zeebaree, S. R., Ahmed, D. M., Sami, A. S., & Zebari, R. R. (2021). IoT and Cloud computing issues, challenges and opportunities: A review. *Qubahan Academic Journal*, *1*(2), 1-7.

Salah, K., Rehman, M. H. U., Nizamuddin, N., & Al-Fuqaha, A. (2019). Blockchain for AI: Review and open research challenges. *IEEE Access*, *7*, 10127-10149.

Shah, J. L., & Bhat, H. F. (2020). CloudIoT for Smart Healthcare: Architecture, Issues, and Challenges. *Internet of Things Use Cases for the Healthcare Industry*, 87-126.

Shao, X. F., Liu, W., Li, Y., Chaudhry, H. R., & Yue, X. G. (2021). Multistage implementation framework for smart supply chain management under industry 0.4. *Technological Forecasting and Social Change*, *162*, 120354.

Shao, X. F., Liu, W., Li, Y., Chaudhry, H. R., & Yue, X. G. (2021). Multistage implementation framework for smart supply chain management under industry 0.4. *Technological Forecasting and Social Change*, *162*, 120354.

Sharma, M., Kamble, S., Mani, V., Sehrawat, R., Belhadi, A., & Sharma, V. (2021). Industry 0.4 adoption for sustainability in multi-tier manufacturing supply chain in emerging economies. *Journal of cleaner production*, *281*, 125013.

Silva, B. N., Khan, M., & Han, K. (2018). Towards sustainable smart cities: A review of trends, architectures, components, and open challenges in smart cities. *Sustainable Cities and Society*, *38*, 697-713.

Sony, M., & Naik, S. S. (2019). Ten lessons for managers while implementing Industry 0,4. *IEEE Engineering Management Review*, *47*(2), 45-52.

Umar, M., Khan, S. A. R., Yusliza, M. Y., Ali, S., & Yu, Z. (2021). Industry 0,4 and green supply chain practices: an empirical study. *International Journal of Productivity and Performance Management*.

Wang, D., Zhong, D., & Souri, A. (2021). Energy management solutions in the Internet of Things applications: Technical analysis and new research directions. *Cognitive Systems Research*, *67*, 33-49.

Wu, L., Yue, X., Jin, A., & Yen, D. C. (2016). Smart supply chain management: a review and implications for future research. *The International Journal of Logistics Management*.

Yadav, G., Luthra, S., Jakhar, S. K., Mangla, S. K., & Rai, D. P. (2020). A framework to overcome sustainable supply chain challenges through solution measures of industry 0,4 and circular economy: An automotive case. *Journal of Cleaner Production*, *254*, 120112.

Yi, K. J., & Jeong, Y. S. (2021). Smart factory: security issues, challenges, and solutions. *Journal of Ambient Intelligence and Humanized Computing*, 1-14.

.

# Quantitative Analysis of Implementation Challenges of IoT-Based Digital Supply Chain (Supply Chain 0/4)

## Hamed Nozari[1*],mohammad Ebrahim sadeghi[2],Javid Ghahremani nahr[3],Seyyed Esmaeil Najafi [4]

## Abstract

Over the past thirty years, logistics has undergone tremendous change from a purely operational performance that led to sales or production and focused on securing supply and delivery lines to customers, with the transformation of intelligent technologies into a professional operator. In today's world, the fourth generation of industry has forced companies to rethink how their supply chain is designed. In this case, in addition to the need for adaptation, supply chains also have the opportunity to reach operational horizons, use emerging digital supply chain business models, and transform the company into a digital supply chain. One of the transformational technologies that has had a tremendous impact on the supply chain in this regard is IoT technology. This technology, as one of the largest sources of data production, can facilitate supply chain processes in all its dimensions. However, due to the presence of the Internet and the location of supply chain components in the context of information networks, this digital supply chain always faces major challenges. Therefore, in this paper, an attempt was made to examine and prioritize the most important challenges of implementing a supply chain 0/4 using a nonlinear hierarchical analysis method. In order to investigate these challenges, the opinions of experts active in the supply chain of Fast-moving consumer goods industries (FMCG) were used as a case study as well as some academic experts.The results show that the lack of technological infrastructure and security challenges are the most important challenges of implementing supply chain 0/4 in the era of digital developments, which should be given special attention for a successful implementation.